\documentclass[aps,preprint,superscriptaddress,titlepage,12pt,floatfix,showpacs,tightenlines]{revtex4}

\usepackage{graphicx}
\usepackage{subfigure}

\input{babarsym.tex}
\def\Dztokpi    {\ensuremath{\Dz \to K^{-}\pi^{+}}\xspace}
\def\Dzbtokpi   {\ensuremath{\Dzb \to K^{+}\pi^{-}}\xspace}
\def\DztokpiWS  {\ensuremath{\Dz \to K^{+}\pi^{-}}\xspace}

\newcommand{\kevcc}{\ensuremath{{\mathrm{\,Ke\kern -0.1em V\!/}c^2}}\xspace}

\def\Rws        {\ensuremath{R_{W\!S}}\xspace}

\def\vrws {0.372} \def\verws{0.025}
\def\vesprws{0.009} \def\vesnrws{0.014}

\begin{document}

\vspace*{-3\baselineskip}
\resizebox{!}{3cm}{\includegraphics{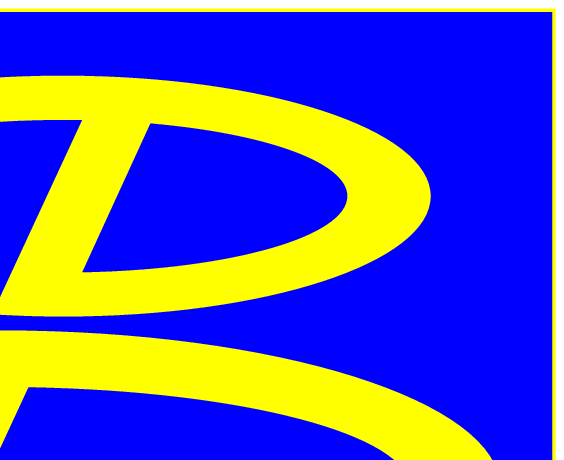}}

\preprint{\vbox{ \hbox{   }
                 \hbox{BELLE-CONF-0254}
                 \hbox{ICHEP02 Parallel 8}
                 \hbox{ABS744}
                 \hbox{hep-ex/0208051}
}}

\title{A measurement of the rate of wrong-sign decays \DztokpiWS}
\affiliation{Aomori University, Aomori}
\affiliation{Budker Institute of Nuclear Physics, Novosibirsk}
\affiliation{Chiba University, Chiba}
\affiliation{Chuo University, Tokyo}
\affiliation{University of Cincinnati, Cincinnati OH}
\affiliation{University of Frankfurt, Frankfurt}
\affiliation{Gyeongsang National University, Chinju}
\affiliation{University of Hawaii, Honolulu HI}
\affiliation{High Energy Accelerator Research Organization (KEK), Tsukuba}
\affiliation{Hiroshima Institute of Technology, Hiroshima}
\affiliation{Institute of High Energy Physics, Chinese Academy of Sciences, Beijing}
\affiliation{Institute of High Energy Physics, Vienna}
\affiliation{Institute for Theoretical and Experimental Physics, Moscow}
\affiliation{J. Stefan Institute, Ljubljana}
\affiliation{Kanagawa University, Yokohama}
\affiliation{Korea University, Seoul}
\affiliation{Kyoto University, Kyoto}
\affiliation{Kyungpook National University, Taegu}
\affiliation{Institut de Physique des Hautes \'Energies, Universit\'e de Lausanne, Lausanne}
\affiliation{University of Ljubljana, Ljubljana}
\affiliation{University of Maribor, Maribor}
\affiliation{University of Melbourne, Victoria}
\affiliation{Nagoya University, Nagoya}
\affiliation{Nara Women's University, Nara}
\affiliation{National Kaohsiung Normal University, Kaohsiung}
\affiliation{National Lien-Ho Institute of Technology, Miao Li}
\affiliation{National Taiwan University, Taipei}
\affiliation{H. Niewodniczanski Institute of Nuclear Physics, Krakow}
\affiliation{Nihon Dental College, Niigata}
\affiliation{Niigata University, Niigata}
\affiliation{Osaka City University, Osaka}
\affiliation{Osaka University, Osaka}
\affiliation{Panjab University, Chandigarh}
\affiliation{Peking University, Beijing}
\affiliation{Princeton University, Princeton NJ}
\affiliation{RIKEN BNL Research Center, Brookhaven NY}
\affiliation{Saga University, Saga}
\affiliation{University of Science and Technology of China, Hefei}
\affiliation{Seoul National University, Seoul}
\affiliation{Sungkyunkwan University, Suwon}
\affiliation{University of Sydney, Sydney NSW}
\affiliation{Tata Institute of Fundamental Research, Bombay}
\affiliation{Toho University, Funabashi}
\affiliation{Tohoku Gakuin University, Tagajo}
\affiliation{Tohoku University, Sendai}
\affiliation{University of Tokyo, Tokyo}
\affiliation{Tokyo Institute of Technology, Tokyo}
\affiliation{Tokyo Metropolitan University, Tokyo}
\affiliation{Tokyo University of Agriculture and Technology, Tokyo}
\affiliation{Toyama National College of Maritime Technology, Toyama}
\affiliation{University of Tsukuba, Tsukuba}
\affiliation{Utkal University, Bhubaneswer}
\affiliation{Virginia Polytechnic Institute and State University, Blacksburg VA}
\affiliation{Yokkaichi University, Yokkaichi}
\affiliation{Yonsei University, Seoul}
  \author{K.~Abe}\affiliation{High Energy Accelerator Research Organization (KEK), Tsukuba} 
  \author{K.~Abe}\affiliation{Tohoku Gakuin University, Tagajo} 
  \author{N.~Abe}\affiliation{Tokyo Institute of Technology, Tokyo} 
  \author{R.~Abe}\affiliation{Niigata University, Niigata} 
  \author{T.~Abe}\affiliation{Tohoku University, Sendai} 
  \author{I.~Adachi}\affiliation{High Energy Accelerator Research Organization (KEK), Tsukuba} 
  \author{Byoung~Sup~Ahn}\affiliation{Korea University, Seoul} 
  \author{H.~Aihara}\affiliation{University of Tokyo, Tokyo} 
  \author{M.~Akatsu}\affiliation{Nagoya University, Nagoya} 
  \author{M.~Asai}\affiliation{Hiroshima Institute of Technology, Hiroshima} 
% ...
\collaboration{The Belle Collaboration}

\begin{abstract}
  A \Dz meson can decay to $K^+\pi^-$ through doubly Cabibbo-suppressed
decay or via  $\Dz\!-\!\Dzb$ mixing.
  With 46.2~fb$^{-1}$ of integrated luminosity collected by Belle,
we have measured the time integrated rate of the wrong-sign process
\DztokpiWS relative to that of the Cabibbo-favored process
\Dztokpi to be $\Rws=\left(\vrws\pm\verws^{+\vesprws}_{-\vesnrws}\right)\%$ (preliminary).
The $\Dz\!-\!\Dzb$ mixing parameters can be derived from the time distribution of the
wrong-sign process.
\end{abstract}

 

\maketitle

\section{Introduction}
The Standard Model predicts that $\Dz\!-\!\Dzb$ mixing is small, with
many estimates for the mixing parameters 
$x = \Delta M/\Gamma$ and $y=\Delta\Gamma/ 2\Gamma$
at the $\lesssim 10^{-3}$ level, although long-distance effects 
in the Standard Model may raise both parameters to 
$\sim10^{-2}$~\cite{Bigi:2000wn,Golowich:1998pz}.
New physics effects may enhance $x$, but are not expected to affect $y$.
 Consequently, a discovery of
large $x$ with small $y$ would imply new physics.

  Experimentally, mixing can be identified by the study of
the wrong-sign (WS) process \DztokpiWS, which can proceed directly 
through doubly  Cabibbo-suppressed decay (DCSD) or through mixing
followed by Cabibbo-favored decay (CFD).
 The decay of an initial $\Dstar{}^+\to\Dz\pi_s^+ $ 
yields a charged ``slow'' pion, denoted $\pi_s^+$, whose sign can be used 
to tag the initial $D$ as either \Dz or \Dzb. \par

The differential WS rate relative to 
the right-sign process is~\cite{ref:wst}
\begin{eqnarray}
r_{WS}(t)
 & = & {\Gamma(\Dz(t)\to K^+\pi^-)
\over \Gamma(\Dztokpi)} \nonumber \\
 & = & \left[R_D + \sqrt{R_D}y'\Gamma t
	+{1\over 4}(x'^2+y'^2)\Gamma^2t^2\right]e^{-\Gamma t},
\end{eqnarray}
where $R_D$ is the relative rate of DCSD to CFD. We use the convention
of CLEO~\cite{ref:cleo}, with  $y' = y\cos\delta - x\sin\delta $ and 
$ x' = x\cos\delta + y\sin\delta $, where $\delta$ is the relative
strong phase between the DCSD and CFD channels.
A fit to the $r_{WS}(t)$ distribution in Belle data, to obtain the mixing
parameters, is underway.  In this paper, we present a preliminary measurement
of the time-integrated wrong-sign rate,
\begin{equation}
\Rws = R_D + \sqrt{R_D}y' + {1\over 2}(x'^2 + y'^2),
\end{equation}
based on the same event reconstruction and background-handling techniques
used for the full analysis.

\section{Experimental Apparatus and Data sample}
  The analysis is based on data accumulated by the Belle detector at
KEKB~\cite{ref:kekb}, an asymmetric collider with 8 GeV electron 
and 3.5 GeV positron storage rings. The data set corresponds to 
an integrated luminosity of 46.2~fb$^{-1}$.

  Belle~\cite{ref:belle}
 is a general-purpose detector based on a 1.5 T superconducting
solenoid. Charged tracks are reconstructed using a 50-layer Central Drift
Chamber (CDC) and a 3-layer double sided Silicon Vertex Detector (SVD).
Particle identification for charged kaons and pions is performed by
combining information from the CDC ($dE/dx$), a set of time-of-flight 
counters (TOF) and a set of aerogel \v Cerenkov counters (ACC).
The combined result of the three systems provides 
$K/\pi$ separation up to 3.5 GeV.

\section{Candidate Reconstruction}
  For \DztokpiWS decay, we use oppositely charged tracks where
both tracks have good SVD hits.
Charged $K$ and $\pi$ mesons are required to be positively identified,
with efficiencies 88.0\% and 88.5\% respectively, and fake rates of 
8.5\% ($\pi$ fakes $K$) and 8.8\% ($K$ fakes $\pi$).
$\Dstar{}^+$ candidates are then reconstructed by combining the \Dz
candidate with another charged track, $\pi_s^+$.
The reconstructed $\Dstar{}^+$ momentum 
in the center of mass frame $P^*(\Dstar)$ is required to be greater than
2.5 GeV to eliminate $B\bar B$ events. Vertex fits are performed,
requiring the two tracks forming the \Dz candidate to originate
from a common vertex, the \Dz flight direction to be consistent with
a particle originating from the interaction point, and the $\pi_s^+$ to
originate from the \Dz production vertex.
We accept candidates with good vertex quality.
The value of \Dz mass $M$ and the energy released in $\Dstar{}^+$ decay
$Q=M(K^+,\pi^-,\pi_s^+) - M(K^+,\pi^-) - M(\pi_s^+)$ are then obtained.

  The right-sign decay $\Dstar{}^+\to\Dz\pi_s^+ $, \Dztokpi is
reconstructed in the same way. Here and throughout this paper, the
inclusion of charge conjugate modes is implied.

\section{Background Determination}
The information in the signal region alone is insufficient to determine the
signal time distribution, because
the background level is comparable to the signal in the signal box,
and the various backgrounds have different time distributions.
We therefore divide the background into categories, and
determine the population of each using $M, Q$ information.
To obtain the signal yield as well as the level of each background,
we perform a two-dimensional (2D) fit to $(M, Q)$.
By studying a large number of continuum Monte-Carlo (MC) events,
we categorize the backgrounds as follows:
\begin{itemize}
\item  $\Dz\to K^-\pi^+$ with double misidentification ($K^-\to\pi^-,\pi^+\to K^+$).
\item  $\Dz\to K^+K^-$  with  single misidentification ($K^-\to\pi^-$).
\item  $\Dz\to \pi^+\pi^-$  with single misidentification ($\pi^+\to K^+$).
\item  \Dzbtokpi combines with a random slow $\pi^+$ to form
a fake $\Dstar{}^+$.
\item  $\Dz\ge3$ body decay, with two daughters identified as $K,\pi$.
\item  $D^+ , \Ds$ decay, with two daughters identified as $K,\pi$.
\item Pure combinatorial background.
\end{itemize}

The $\Dz\to K^-\pi^+$ double misidentification background
can be rejected by kinematic requirements.
We evaluate the \Dz mass under swapped mass assignment 
\(M_{\text{flip}}=M_{K\to\pi,\pi\to K}\). If $M_{\text{flip}}$ falls
within 28 MeV ($\sim4\sigma$) of the nominal \Dz mass, the \DztokpiWS
candidate is rejected.
For the $\Dz\to K^+K^-,\pi^+\pi^-$ single misidentification background,
the change from the hypothesis that a charged track is a $\pi(K)$
to a $K(\pi)$ in the \Dz 2 body decay daughters results in
an increase(decrease) in the reconstructed \Dz mass $M$ of
at least 60 MeV. This behavior has been studied in both data and MC.
So we choose the \Dz mass window $1.81\sim1.91$ GeV for histogram
making and fitting to exclude these two backgrounds.

\begin{figure}[p]
\includegraphics[width=0.5\textwidth]{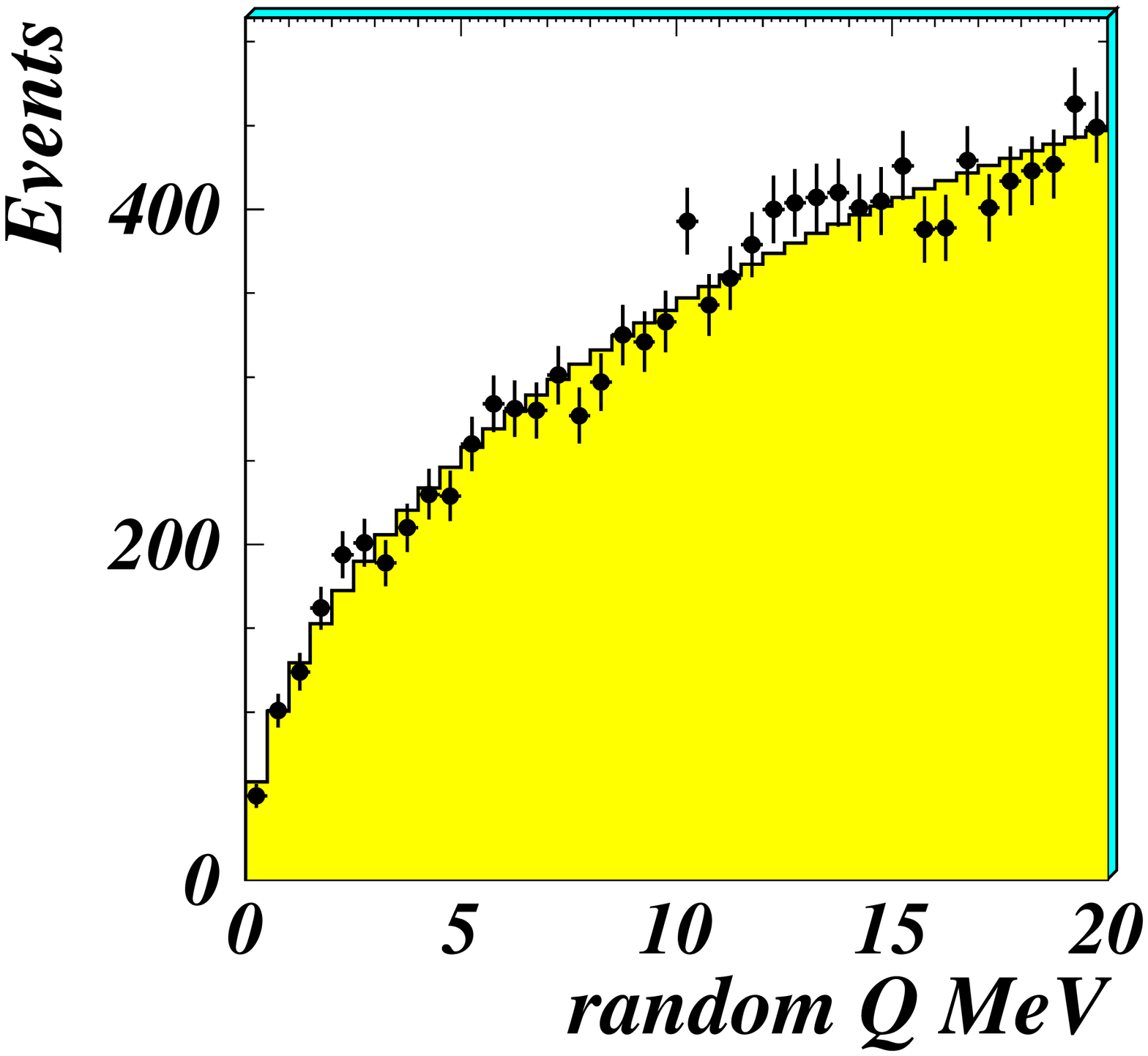}
\caption{\label{fig:rnd} The $Q$ distribution for MC events (points)
and the fit function (histogram) for
the random slow $\pi^+$ background. The $M$ shape is
fixed to that of the right-sign \Dz signal in the data.}
\vspace*{4ex}
\includegraphics[width=0.5\textwidth]{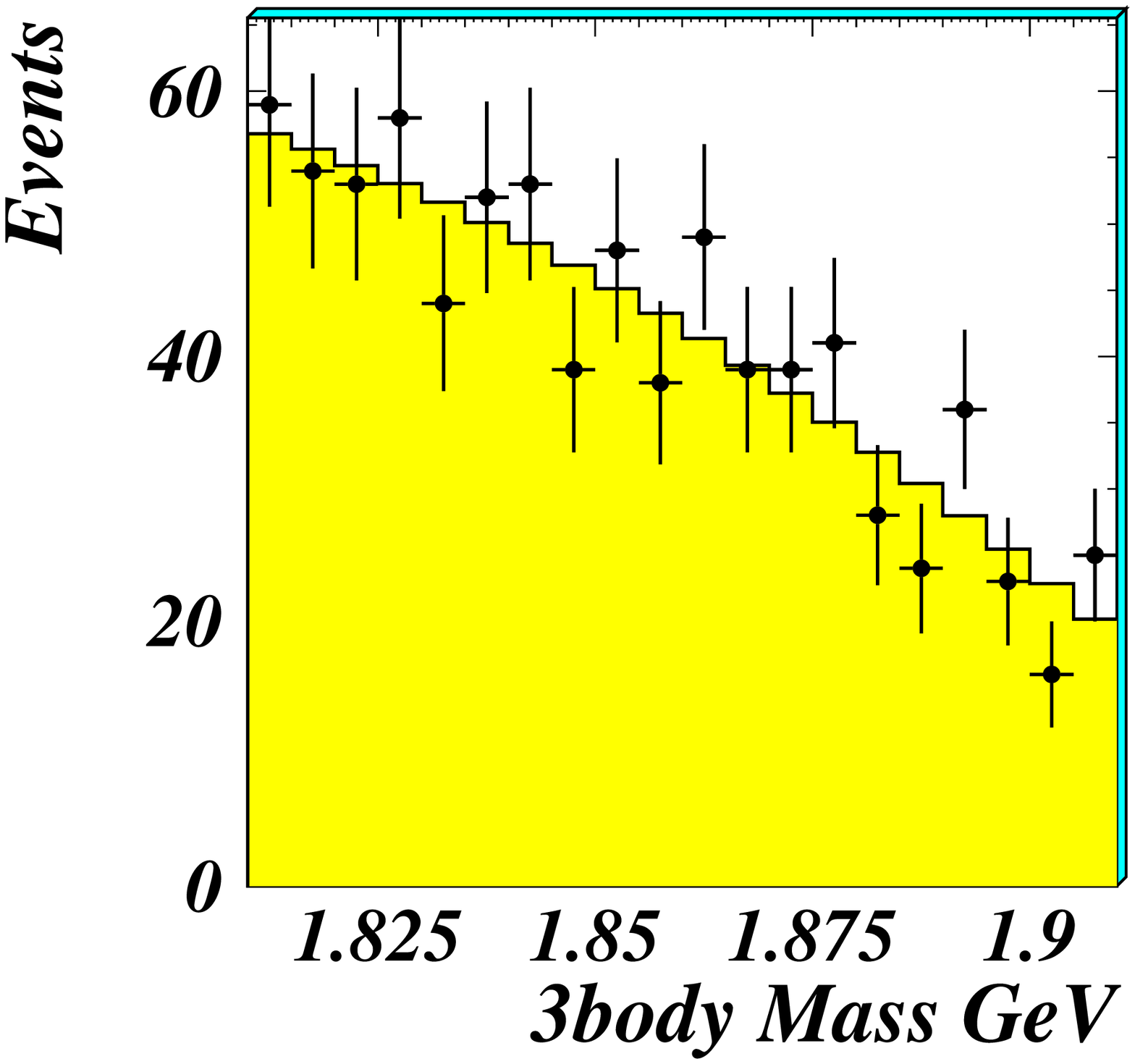}%
\includegraphics[width=0.5\textwidth]{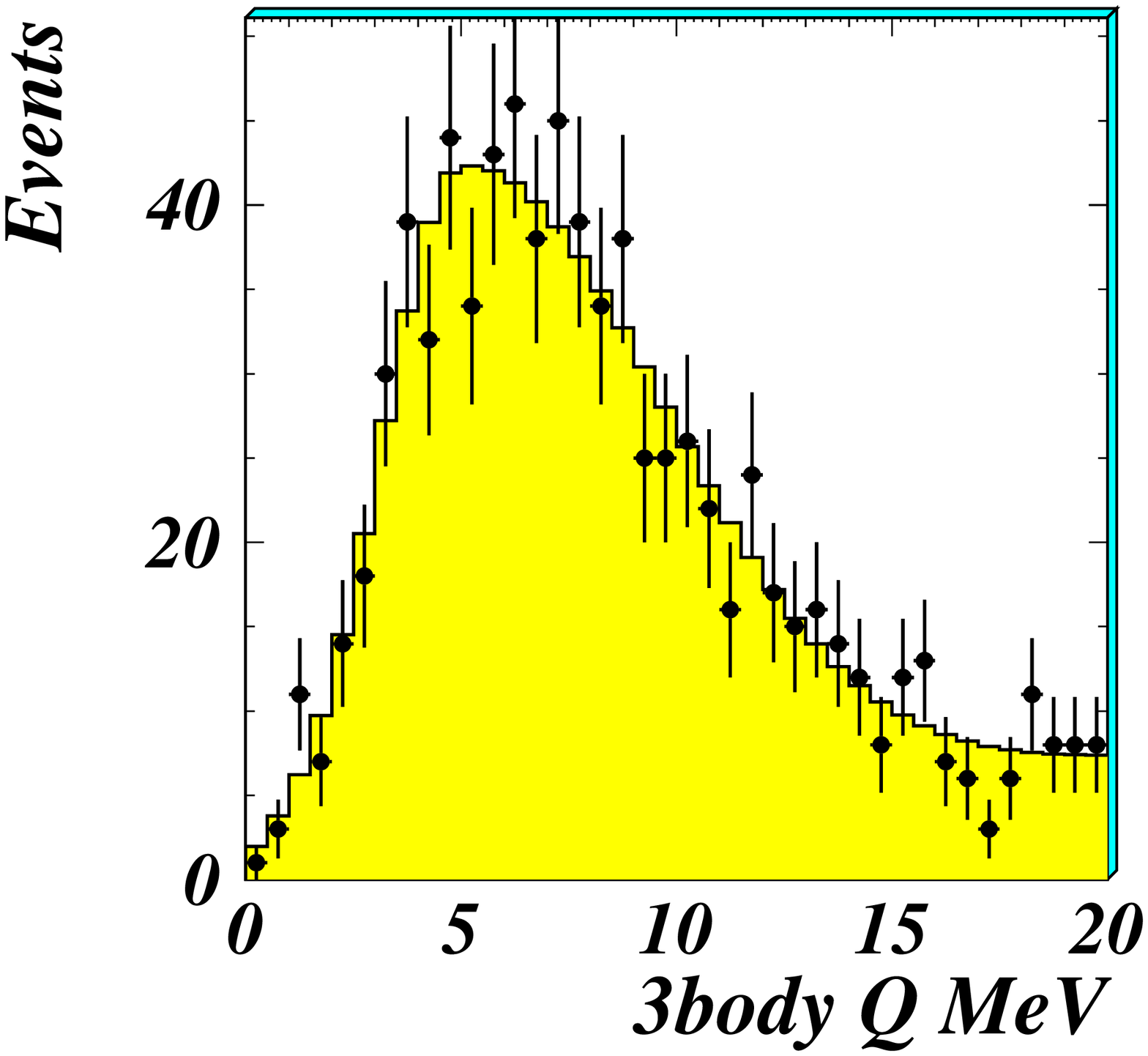}%
\caption{\label{fig:d3b}  $M$ (left) and $Q$ (right)
 distributions for MC events (points)
and the fit functions (histograms) for the $\Dz\ge3$ body background.}
\end{figure}

\begin{figure}[p]
\includegraphics[width=0.5\textwidth]{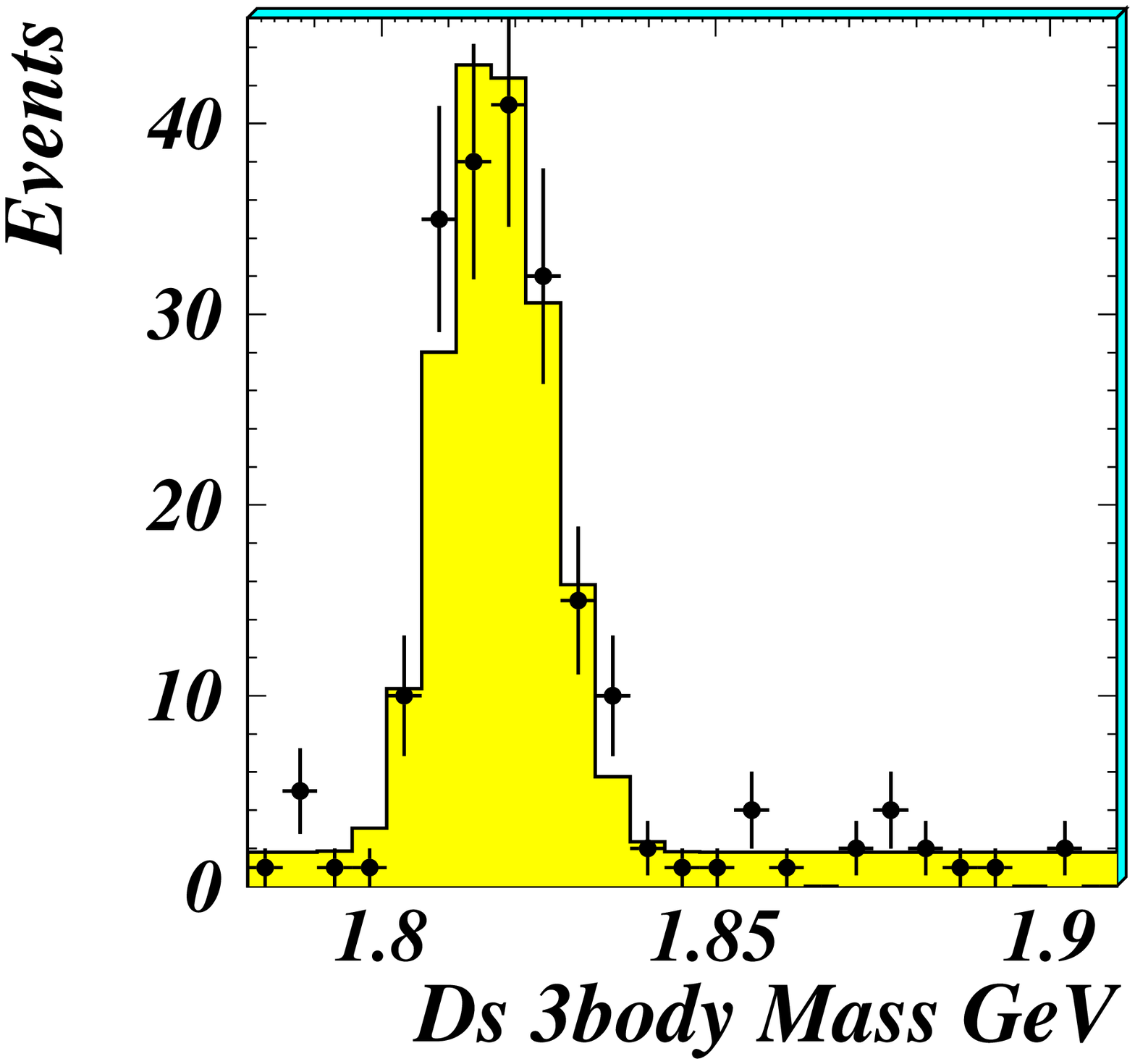}%
\includegraphics[width=0.5\textwidth]{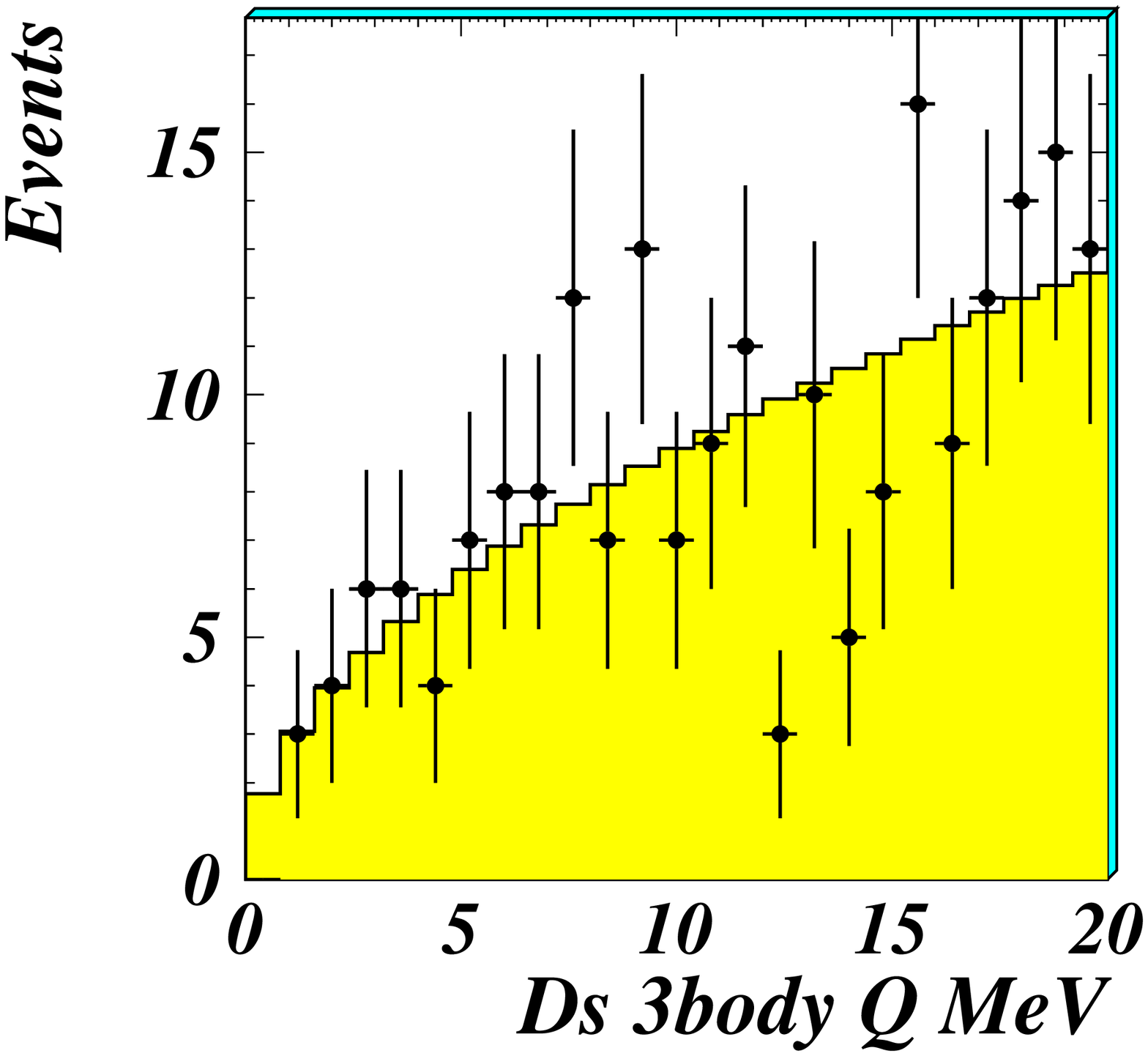}%
\caption{\label{fig:ds3}  $M$ (left) and $Q$ (right)
distributions for MC events (points)
and the fit functions (histograms) for the $\Ds,D^+$ background.}
\vspace*{4ex}
\includegraphics[width=0.5\textwidth]{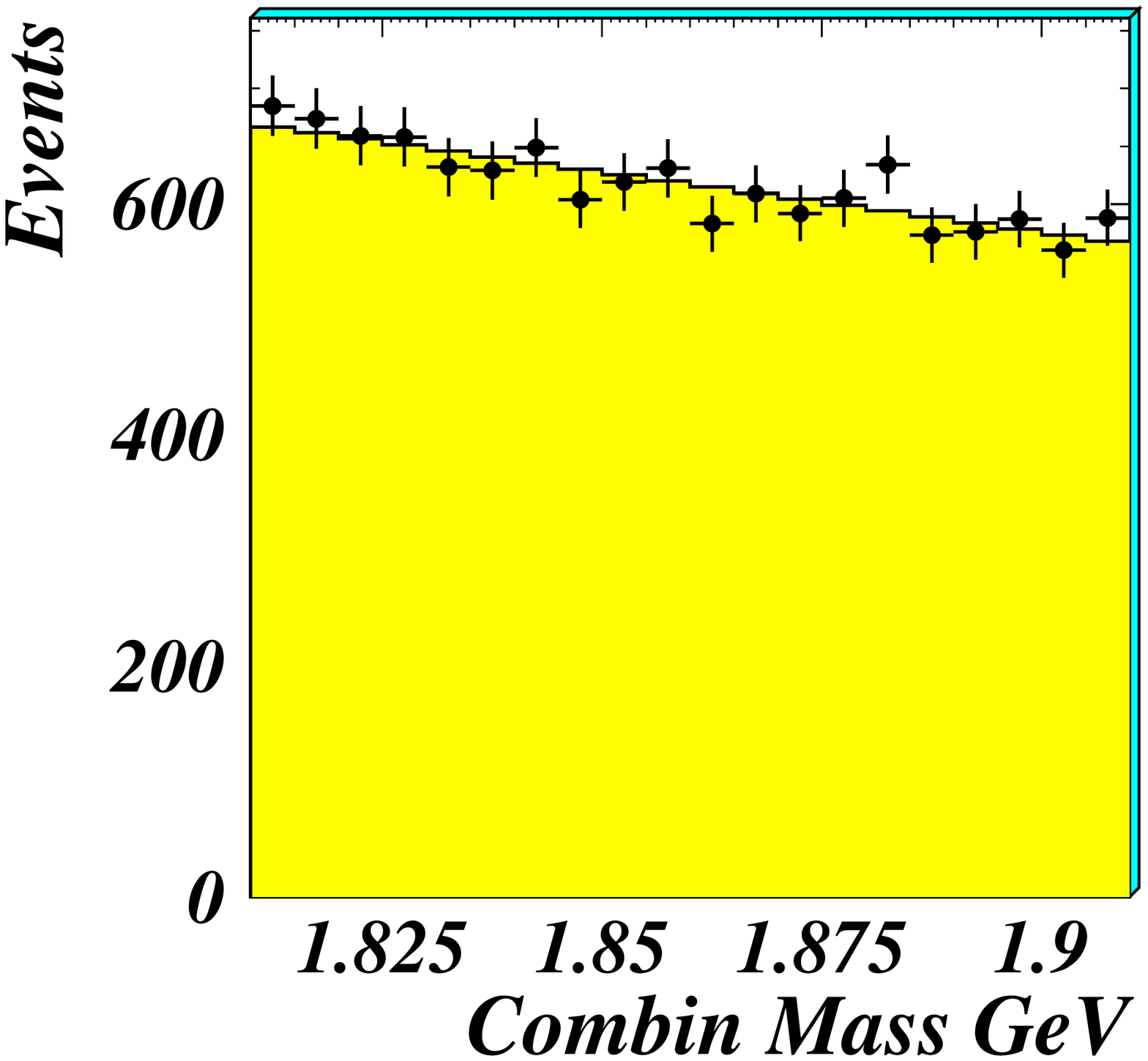}%
\includegraphics[width=0.5\textwidth]{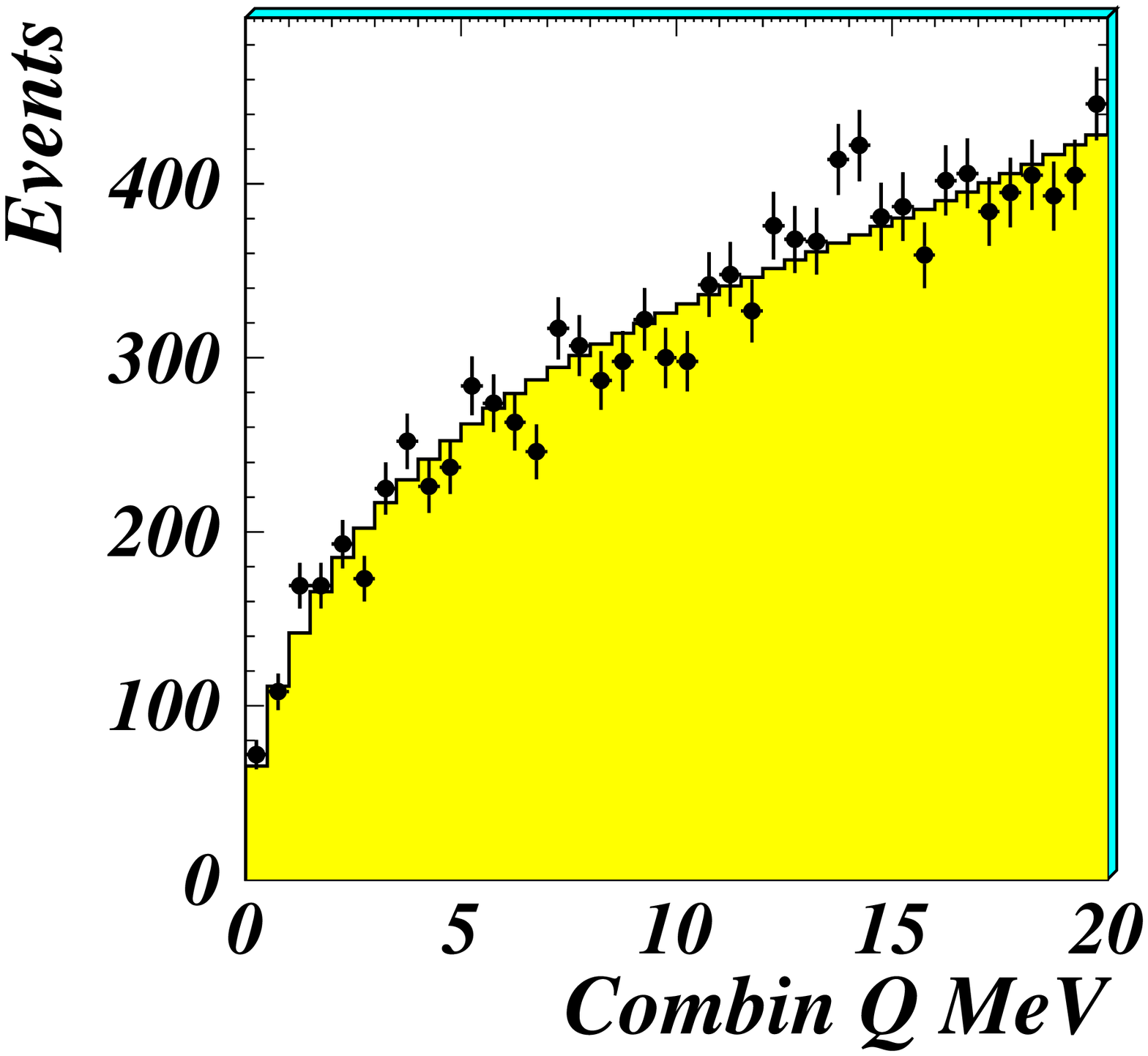}%
\caption{\label{fig:cmb}  $M$ (left) and $Q$ (right)
distributions for MC events (points)
and the fit functions (histograms) for the combinatorial background.}
\end{figure}

   Each of the remaining four types of background is 
parametrized and fitted with a 2D function to obtain its shape,
as shown in Figs~\ref{fig:rnd}--\ref{fig:cmb}.
  The backgrounds in the right-sign sample are categorized and fitted 
in a similar way. They are grouped into three types:
\begin{itemize}
\item  \Dztokpi combines with a random slow $\pi^+$ to form
a fake $\Dstar{}^+$.
\item  Charmed ground state meson $\ge3$ body decay,
 with two daughters identified as $K,\pi$ (``$D\ge$3 body'' background).
\item Pure combinatorial background.
\end{itemize}

\section{Results}
\begin{figure}[htbp]
\includegraphics[width=0.5\textwidth]{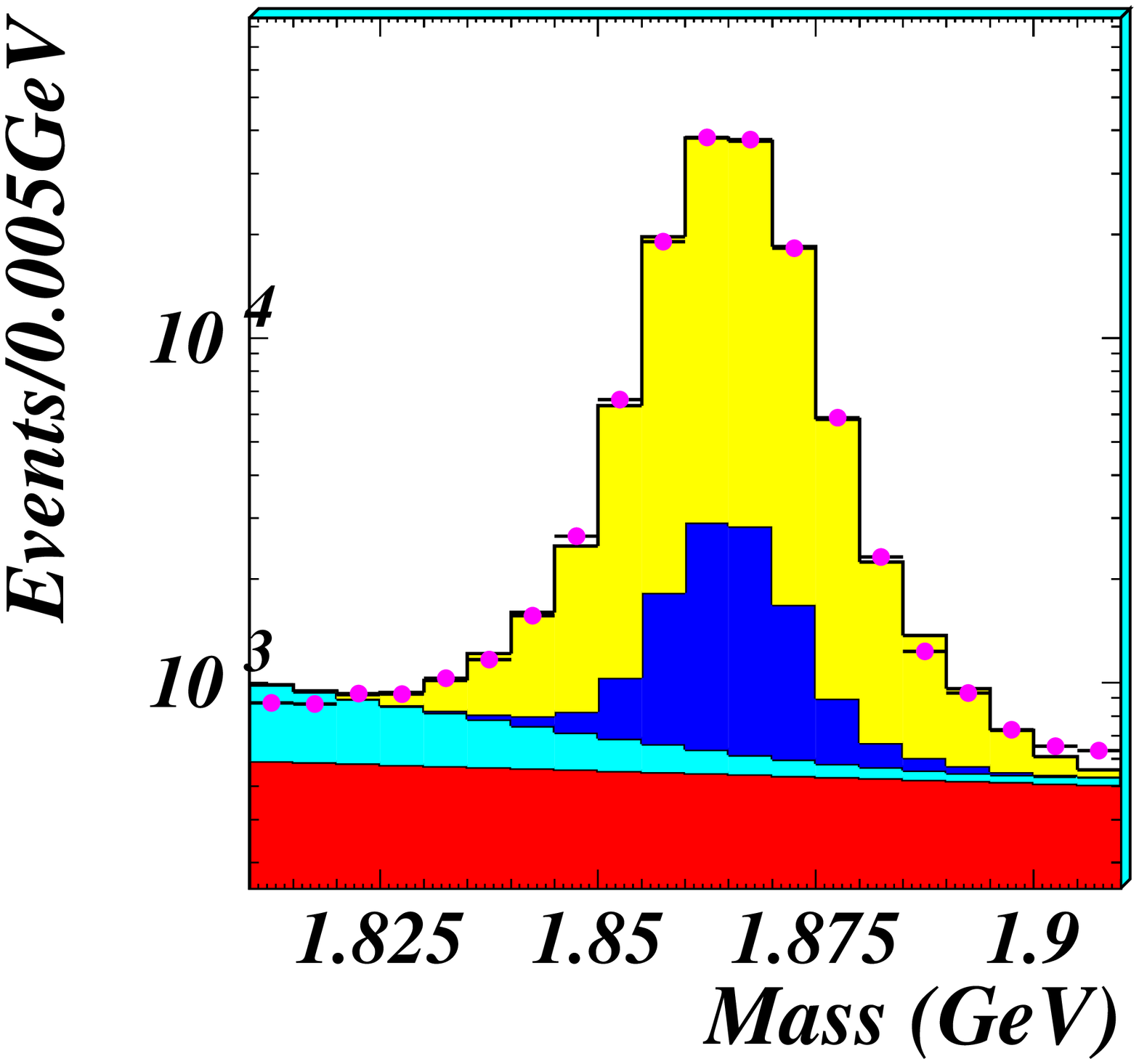}%
\includegraphics[width=0.5\textwidth]{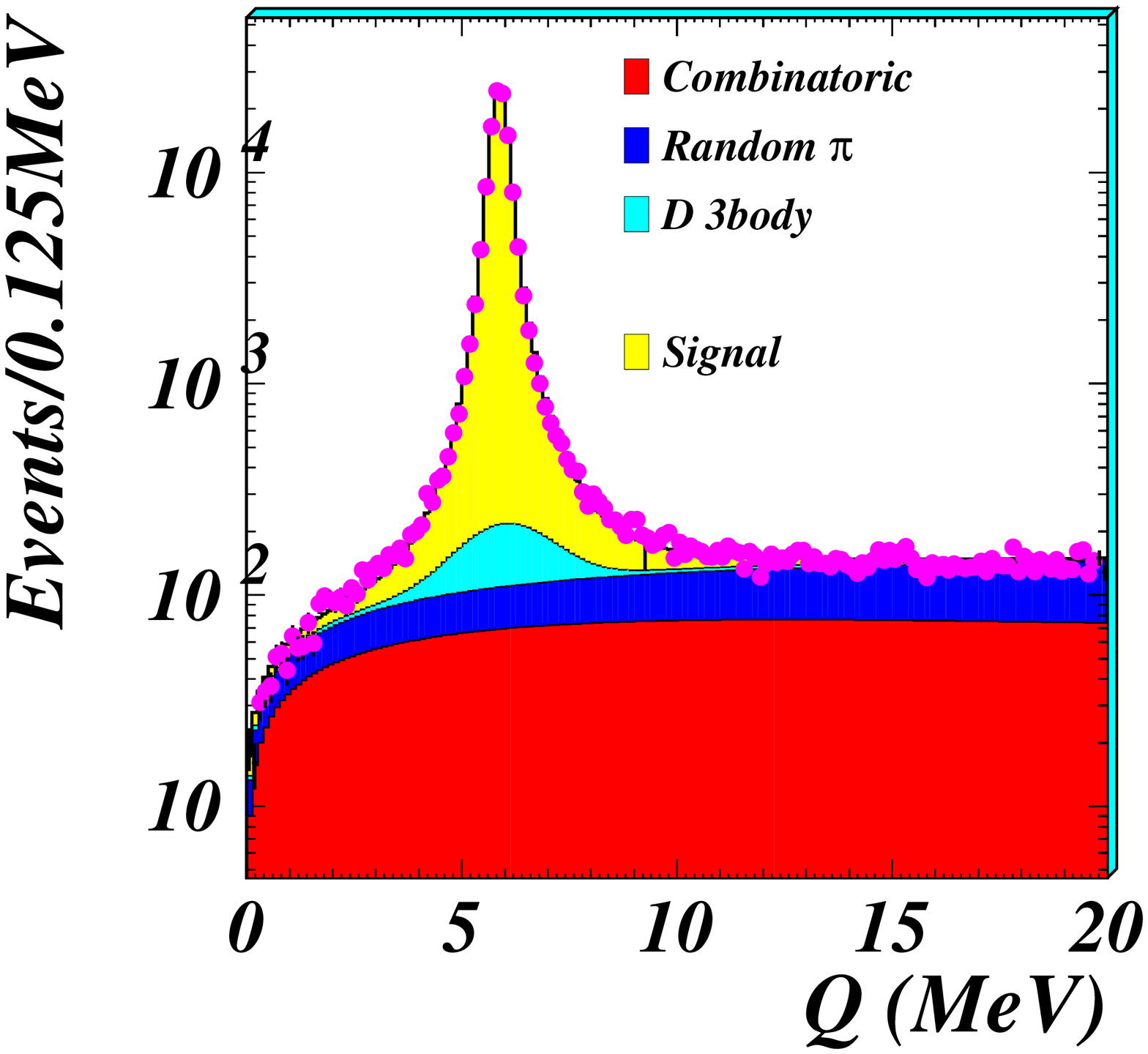}%
\caption{\label{fig:rs}  Projections of $M$ (left) and $Q$ (right)
for the right-sign data (points) and the fit functions (histograms),
for the region $1.81\le M<1.91$ GeV and $0\le Q<20$ MeV.
Note the logarithmic scale.}
\end{figure}

\begin{figure}[htbp]
\includegraphics[width=0.5\textwidth]{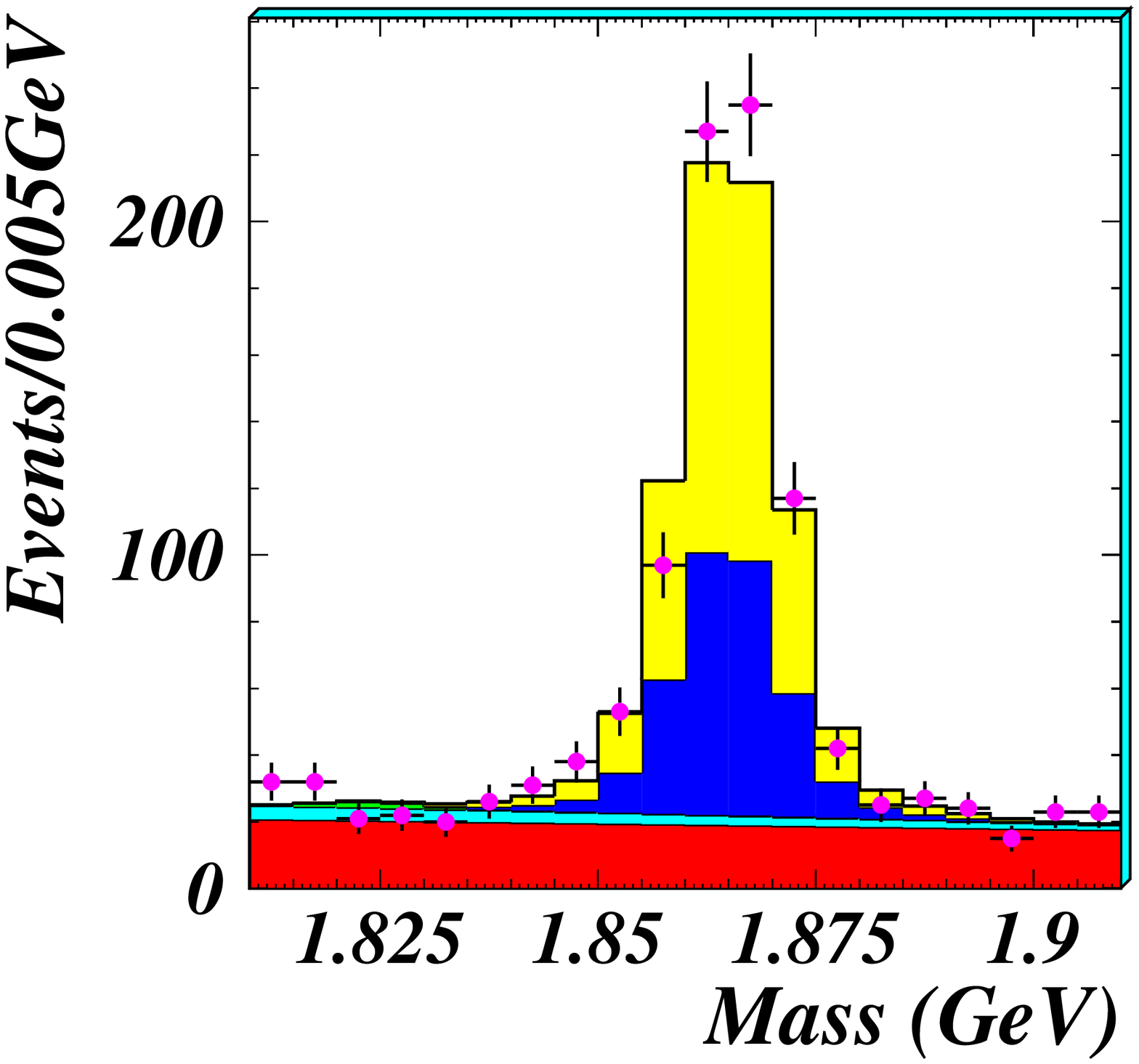}%
\includegraphics[width=0.5\textwidth]{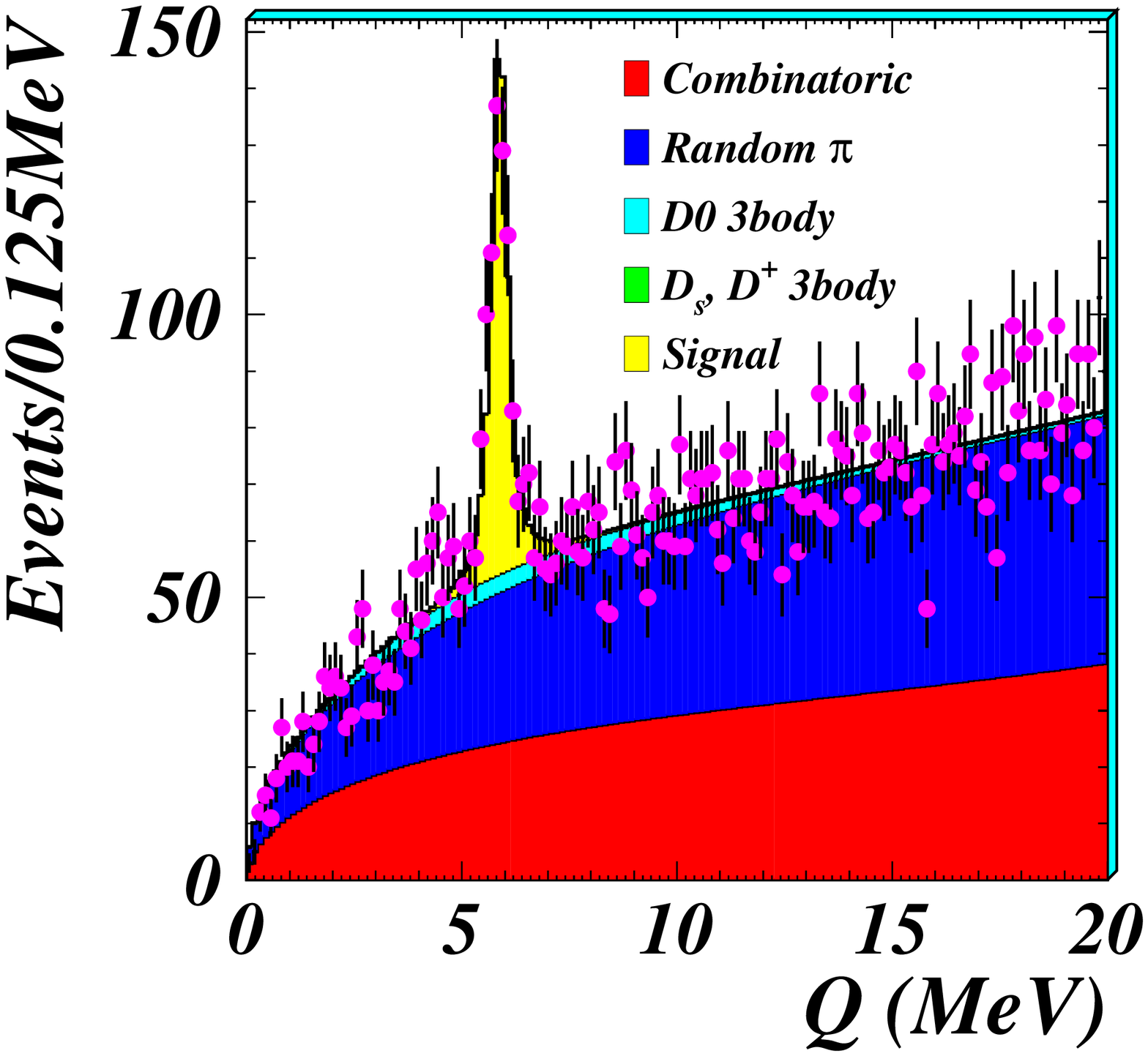}%
\caption{\label{fig:wss}  Projections of $M$ (left) and $Q$ (right)
for the wrong-sign data (points) and the fit functions (histograms), within
a $3\sigma$ window in the complementary variable
($5.27\le Q<6.47$ MeV and $1.8445\le M<1.8845$ GeV respectively).
The signal contribution is shaded yellow.}
\end{figure}

We perform a 2D fit to the ($M,Q$) distribution for the right-sign data,
with the normalization of each background allowed to float in the fit.
The signal is represented
by a double Gaussian in $M$ and a bifurcated Student's t function in $Q$.
The projections onto $M$ and $Q$ are shown in Fig.~\ref{fig:rs}.
To perform the 2D fit
for the wrong-sign data, we fix the wrong-sign signal shape to 
the shape of the right-sign signal, and
float the normalization of the signal and each type of background, but 
fix the \textit{relative} normalization between the
$\Ds,D^+$ and $\Dz\ge3$ body backgrounds.
The projections of the wrong-sign data and the fit results are
shown in Fig.~\ref{fig:wss}. The data are the circles with error bars, and
there are 20 bins in $M$ ($1.81\sim1.91$ GeV)
and 160 bins in $Q$ ($0\sim20$ MeV).
We find $120795\pm 371$ \Dztokpi and $450\pm 31$ \DztokpiWS events.
The ratio \Rws is then calculated to be $\left(\vrws\pm\verws\right)\%$.

  Many systematics cancel because \Rws is a ratio between two
decay modes with similar kinematics. The dominant systematic
errors stem from potential imperfect modeling of the shapes
for our backgrounds. We vary selection criteria ($K$ and $\pi$ identification
cuts, vertexing criteria, the $P^*(\Dstar)$ cut) and use different
background parameterizations, and then repeat the fit on each category of 
MC background as well as on the right-sign and wrong-sign data.
The variation in the result is taken as a measure of the systematic error.
We also vary the background shape parameters according to their errors and
calculate the resulting error on \Rws. The correlations among
these background shape parameters are considered either in the variation or
in the \Rws error calculation.
The results are summarized in Table~\ref{tab:esys}. All systematic errors
are combined in quadrature.
\begin{table}
\caption{\label{tab:esys} Summary of systematic errors on \Rws}
\begin{ruledtabular}
\begin{tabular}{lc}
Source       		&   Systematic error(\%) \\ \hline
Kaon identification	& $^{+0.0067}_{-0.0023}$ \\
Pion identification     & $^{+0.0023}_{-0.0048}$ \\
Vertex fit	 	& $^{+0.0042}_{-0.0051}$ \\
$P^*(\Dstar)$ cutoff	& $^{+0.0000}_{-0.0117}$ \\
Background shapes	& $^{+0.0027}_{-0.0026}$ \\
Background parametrizations& $\pm0.0011$	\\ \hline
Total			& $^{+0.009}_{-0.014}$
\end{tabular}
\end{ruledtabular}
\end{table}

\begin{figure}[htbp]
\includegraphics[width=0.7\textwidth]{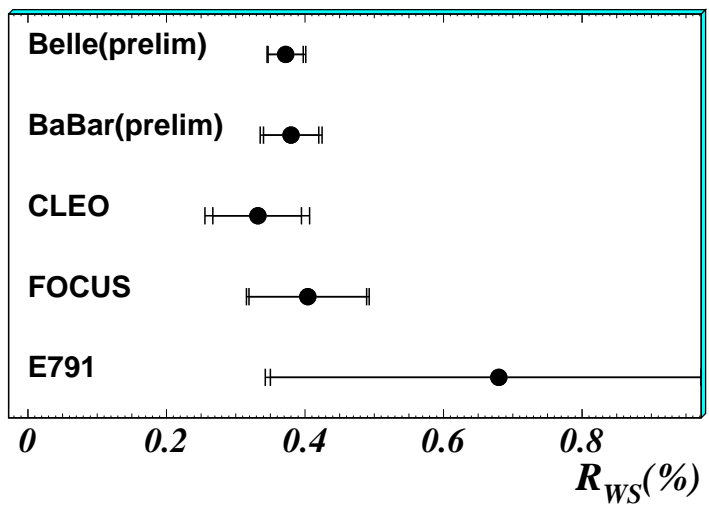}%
\caption{\label{fig:plwor} Various experimental results for \Rws.}
\end{figure}

   In summary, we have measured the ratio \Rws of the rate of the wrong-sign process
\DztokpiWS relative to the right-sign process \Dztokpi to be
\begin{equation}
\Rws = \left( \vrws\pm\verws(\text{stat})^{+\vesprws}_{-\vesnrws}(\text{syst}) \right) \% .
\end{equation}
  This result is preliminary. It is compared with other
experimental results~\cite{ref:babar,ref:cleo,ref:focus,ref:e791}
in Fig~\ref{fig:plwor}.

\section*{Acknowledgments}
We wish to thank the KEKB accelerator group for the excellent
operation of the KEKB accelerator.
We acknowledge support from the Ministry of Education,
Culture, Sports, Science, and Technology of Japan
and the Japan Society for the Promotion of Science;
the Australian Research Council
and the Australian Department of Industry, Science and Resources;
the National Science Foundation of China under contract No.~10175071;
the Department of Science and Technology of India;
the BK21 program of the Ministry of Education of Korea
and the CHEP SRC program of the Korea Science and Engineering Foundation;
the Polish State Committee for Scientific Research
under contract No.~2P03B 17017;
the Ministry of Science and Technology of the Russian Federation;
the Ministry of Education, Science and Sport of the Republic of Slovenia;
the National Science Council and the Ministry of Education of Taiwan;
and the U.S.\ Department of Energy.


\begin{thebibliography}{99}
\bibitem{Bigi:2000wn}
I.~I.~Bigi and N.~G.~Uraltsev,
Nucl.\ Phys.\ B {\bf 592}, 92 (2001).
\bibitem{Golowich:1998pz}
E.~Golowich and A.~A.~Petrov,
Phys.\ Lett.\ B {\bf 427}, 172 (1998).
\bibitem{ref:wst} G.~Blaylock, A.~Seiden, and Y.~Nir,
Phys. Lett. {\bf B 355}, 555 (1995).
\bibitem{ref:kekb} E.~Kikutani ed., KEK Preprint 2001-157 (2001), 
 to appear in Nucl. Instr. and Meth. A.
\bibitem{ref:belle} S.~Mori ed., A.~Abashian {\it et al.}
 (Belle Collaboration),
Nucl. Instr. and Meth. A {\bf 479}, 117 (2002).
\bibitem{ref:babar} U.~Egede (BaBar Collaboration), \texttt{hep-ex/0111062}.
\bibitem{ref:cleo} CLEO Collaboration (R.~Godang {\it et al.}),
Phys. Rev. Lett. {\bf 84}, 5038 (2000).
\bibitem{ref:focus} FOCUS Collaboration (J.M.~Link {\it et al.}),
Phys. Rev. Lett. {\bf 86}, 2955 (2001).
\bibitem{ref:e791} E791 Collaboration (E.M.~Aitala {\it et al.}),
Phys. Rev. {\bf D57}, 13 (1998).
\end{thebibliography}
\end{document}